\documentclass{owrart}

\newcommand\ve{\varepsilon}

\newcommand\lang{\langle\langle}
\newcommand\rang{\rangle\rangle}

\begin{document}


\begin{talk}[Pablo Ferrari]{Stefano Olla}
{Diffusive Fluctuations in Hard Rods System}
{Olla, Stefano}

\noindent
Consider a system of one dimensional hard rods of variable length
  in the dynamics considered in \cite{FGS22}: when two rods collide
  they exchange positions. The equilibrium dynamics is constructed as follows.
  We start with $X^\ve = (x,v,r)$ the Poisson process on $\mathbb R\times \mathbb R \times
  \mathbb R_+$ with intensity $\ve^{-1} \rho\; dx \;d\mu(v,r)$, with $\mu$
a positive probability measure on $\mathbb R^2$ with finite second moments.
The usual hard rods case is given by  $d\mu(v,r) = \delta_{a}(dr) d\mu(v)$, $a>0$.
We define
\begin{equation}
  \label{eq:1}
  \begin{split}
    \sigma &= \rho \iint r d\mu(v,r), \qquad \text{volume density}\\
    \pi &= \rho \iint r v d\mu(v,r), \qquad \text{momentum density}.
\end{split}
\end{equation}
and
\begin{equation}
  \label{eq:5}
  \begin{split}
    m_a^b(X^\ve) =\begin{cases}
      \sum_{(x,v,r)\in X^\ve, x\in [a,b]} \ve r \qquad b>a\\
      - \sum_{(x,v,r)\in X^\ve, x\in [b,a]} \ve r \qquad b < a.
    \end{cases}
\end{split}
\end{equation}
By the law of large numbers we have
\begin{equation}
  m_a^b(X^\ve) 
  \ \mathop{\longrightarrow}_{\ve \to 0}
  (b-a) \sigma, \qquad  \text{a.s. as} \label{eq:7}
\end{equation}
To each configuration $X^\ve$ there is a dilated configuration of the rods
$$
Y^\ve = \{(y = x + m_0^x(X^\ve), v, r): (x,v,r)\in X^\ve\}
$$
For a given test function $\varphi(y,v,r)$ we have the law of large numbers
\begin{equation}
  \label{eq:ylnn}
  \begin{split}
    \ve \sum_{(y,v,r)\in Y^\ve} r\varphi(y, v, r) =
    \ve \sum_{(x,v,r)\in X^\ve} r\varphi(x + m_0^x(X^\ve), v, r)\\
    \ \mathop{\longrightarrow}_{\ve \to 0}\rho \iiint r\varphi(x (1+  \sigma),v,r)
    dx\; d\mu(v,r)\\
    = \frac{\rho}{1 + \sigma}  \iiint r\varphi(y,v,r) dy\; d\mu(v,r)
    = \frac{1}{1+\sigma} \lang \varphi\rang.
  \end{split}
\end{equation}
i.e. $\bar \rho= \frac{\rho}{1 + \sigma}$ is the density of the hard rods.

The fluctuation field for the rods is defined by
\begin{equation}
  \label{eq:73}
  \xi^{Y,\ve}(\varphi) =
  \ve^{-1/2}\left[\ve \sum_{(y,v,r)\in Y^\ve} r\varphi(y, v, r) -
    \mathbb E\left( \ve \sum_{(y,v,r)\in Y^\ve}
        r\varphi(y, v, r)\right) \right].
  \end{equation}

  It is not hard to prove, using the underlying CLT of the Poisson process,
  that
  \begin{equation}
  \label{eq:17}
  \xi^{Y,\ve}(\varphi)  \ \mathop{\longrightarrow}_{\ve \to 0}^{\text{law}}\
  \xi^Y(\varphi) 
\end{equation}
where $\xi^Y$ is the centered gaussian field with covariance
\begin{equation}
  \label{eq:18}
  \begin{split}
    < \xi^Y(\varphi) \xi^Y(\psi) > 
  = \bar \rho \iiint r^2 C\varphi(y,v,r) C\psi(y,v,r) dy d\mu(v,r).
\end{split}
\end{equation}
where $C = I- \frac {\sigma}{1+\sigma} P$ and 
$P$ is the \emph{projection} operator
\begin{equation}
  \label{eq:15}
  P\varphi (x) = \frac{\rho}{\sigma} \iint r \varphi(x, v', r') \; d\mu(v',r').
\end{equation}

In the dynamics that we consider, in the Euler scaling, the position at time $t$
of the hard rod $(y,v,r)$ that at
initial time is at the position
$y$ that is the dilated image (wrt $0$) of the point $x$ is given by
\begin{equation}
   y_t = 
   x +m_0^x(X^\ve) + vt + j_{X^\ve}(x,v,t).
   \label{eq:45}
\end{equation}
The flux $ j_{X^\ve}(x,v,t)$ is defined by
\begin{equation}
  \label{eq:29}
  \begin{split}
    j_{X^\ve}(x,v,t) = \ve\sum_{(x',v',r')\in X^\ve} r'
    \left(1_{[v'<v]} 1_{[x< x'<x+(v-v')t]} - 1_{[v'>v]}1_{[x+(v-v')t<x'<x]}\right).
  \end{split}
\end{equation}

By the law of large numbers, for any tagged rod $(y,v,r) \in Y^{\ve}$
we have that
 \begin{equation}
   \label{eq:22}
    \left(y_{t} - y\right)  \ \mathop{\longrightarrow}_{\ve \to 0} 
    v^{\text{eff}}(v) t,\qquad \text{a.s.}
 \end{equation}
 where the effective velocity is given by
 \begin{equation}
   \label{eq:23}
   v^{\text{eff}}(v) : = v + {\rho} \iint r (v - w) d\mu(w,r)
   = v(1+\sigma) - \pi.
 \end{equation}

 The fluctuation field in the Euler scaling is defined by
 \begin{equation}
   \label{eq:76}
   \begin{split}
     \xi^{Y,\ve}_t (\varphi) =
      \ve^{-1/2}\left[\ve \sum_{(y,v,r)\in Y^\ve}
        r\varphi( y_{t}, v, r) - \frac 1{1+\sigma}  \lang \varphi \rang\right].
   \end{split}
 \end{equation}
 It is proven in \cite{BW88} that $ \xi^{Y,\ve}_t $ converges in law to
 \begin{equation}
   \label{eq:48}
   \xi^Y_t(\varphi) = \xi^Y_0(\varphi_t), \qquad \varphi_t(y,v,r) = \varphi(y+v^{\text{eff}}t,v,r). 
 \end{equation}
i.e.
\begin{equation}
  \label{eq:49}
  \partial_t \xi^Y_t(\varphi) = \xi^Y_0(v^{\text{eff}}\partial_x\varphi_t) =
  \xi^Y_t(v^{\text{eff}}\partial_x\varphi),
\end{equation}
that in the hard rods with deterministic
length correspond to the linerized equation of the
Euler hydrodynamics proven in \cite{BDS80,BDS83}.
Our work concerns the fluctuation field recentered on the effective velocities
under a diffusive rescaling:
\begin{equation}
   \label{eq:21}
   \begin{split}
     \Xi^{Y,\ve}_t (\varphi) =  \ve^{-1/2}\left[\ve \sum_{(y,v,r)\in Y^\ve}
       r\varphi\left[ y_{\ve^{-1}t} - v^{\text{eff}}(v) \ve^{-1} t , v, r\right]-
       \frac 1{1+\sigma} \lang \varphi \rang\right].
   \end{split}
 \end{equation}
We prove the following convergence in law:
 \begin{equation}
   \label{eq:32}
   \Xi^{Y,\ve}_t (\varphi) \mathop{\longrightarrow}_{\ve\to 0}^{\text{law}} \  \Xi^{Y}_t (\varphi)
   = \Xi^{Y} \left(\varphi(\cdot + \sqrt{\mathcal{D}} W_t)\right).
 \end{equation}
 This means that an initial fluctuation of rods of velocity $v$,
 after \emph{recentering}
 around the effective velocity, evolves in the diffusive scale by random
 rigid translations driven by a Brownian motion with diffusivity $\mathcal{D}(v)$
 explicitely defined by
 \begin{equation}
   \label{eq:2}
   \mathcal{D}(v) = \rho\iint  r^2 |v-\bar v| d\mu(\bar v,r).
 \end{equation}
 In the case of the usual hard rods with deterministic length this coincide with
 the diffusivity that appears in the Navier-Stokes corrections of the hydrodynamics
 \cite{BDS90,BS97} (see also more recent \cite{DS-17}).
 This rigidity in the evolution of the fluctuations in the diffusive scaling is in contrast with
 expected results for chaotic systems where fluctuation hydrodynamics predict an
 evolution driven by an additive space-time white noise \cite{SO-Spohn-book}.
 On the other hand, in the case of the ususal hard rods with fixed size,
 it is in agreement with previous calculations of the space-time covariance
 \cite{LPS68,Spohn82}.
 We expect this rigid evolution of the fluctuations in
 other completely integrable system such as the Ball-Box dynamics \cite{CS-22,FG-22}
 or the Toda lattice \cite{S-Toda-21}.

 The basic argument behind the proof of  \eqref{eq:32} is that the
 fluxes $ j_{X^\ve}(x,v,\ve^{-1}t)$ and $ j_{X^\ve}(\bar x,v,\ve^{-1}t)$
 corresponding to two particles at initial macroscopic distance $x-\bar x$
 are completely correlated in the limit as $\ve\to 0$.
 
 \medskip
 \textbf{Aknowledgments:} We thanks Herbert Spohn for very valuable comments
 on this work.

\end{talk}

\end{document}